\documentclass[jcp,aps,showpacs,supergroupedaddress,epsfig,amsmath,amssymb,eqsecnum,twocolumn]{revtex4}
\usepackage{epsfig,amsmath,amsthm,amssymb,bm,epsf,graphics,psfrag,verbatim,subfigure,framed}
\usepackage[makeroom]{cancel}
\usepackage[usenames,dvipsnames]{color}
\usepackage{empheq}
\usepackage{bbm}
\usepackage{mathrsfs}
\usepackage{amsfonts}
\usepackage{color}
\usepackage{pbox}
\def\hv{{\bf h}}

\newcommand{\be}{\begin{equation}}
\newcommand{\ee}{\end{equation}}
\newcommand{\ba}{\begin{eqnarray}}
\newcommand{\ea}{\end{eqnarray}}
\newcommand{\bw}{\begin{widetext}}
\newcommand{\ew}{\end{widetext}}

\newcommand{\Av}{{\bf{A}}}

\newcommand{\rv}{{\mathbf{r}}}

\newcommand{\qv}{{\bf q}}
\newcommand{\cv}{{\bf c}}
\newcommand{\dv}{{\bf d}}

\begin{document}

\title{Van der Waals torque and force between anisotropic topological insulator slabs}
\author{Bing-Sui Lu$^1$}
\email{binghermes@gmail.com, bslu@ntu.edu.sg}
%\author{Rudolf Podgornik$^{1,2}$}
\affiliation{$^{1}$ Division of Physics and Applied Physics, School of Physical and Mathematical Sciences, Nanyang Technological University, 21 Nanyang Link, 637371 Singapore.
}
%\affiliation{$^{2}$Department of Physics, Faculty of Mathematics and Physics, University of Ljubljana, 1000 Ljubljana, Slovenia.}
\date{\today}
%\pacs{82.70.Dd, 83.80.Hj, 82.45.Gj, 52.25.Kn}

\begin{abstract}
We investigate the character of the van der Waals (vdW) torque and force between two coplanar and dielectrically anisotropic topological insulator (TI) slabs separated by a vacuum gap in the non-retardation regime, where the optic axes of the slabs are each perpendicular to the normal direction to the slab-gap interface and also generally differently oriented from each other. 
We find that in addition to the magnetoelectric coupling strength, the anisotropy can also influence the sign of the vdW force, viz., a repulsive vdW force can become attractive if the anistropy is increased sufficiently. In addition, the vdW force oscillates as a function of the angular difference between the optic axes of the TI slabs, being most repulsive/least attractive (least repulsive/most attractive) for angular differences that are integer (half-integer) multiples of $\pi$. 
Our third finding is that the vdW torque for TI slabs is generally weaker than that for ordinary dielectric slabs. 
Our work provides the first instance in which the vector potential appears in a calculation of the vdW interaction for which the limit is non-retarded or static. 
\end{abstract}

\maketitle

\section{Introduction}

The Casimir effect was first discovered~\cite{casimir} by H. B. G. Casimir in 1948 in the theoretical context of two perfectly conducting surfaces interacting across a vacuum at zero temperature. The effect was then generalized to the case of dielectric slabs interacting across a dielectrically dissimilar gap at a finite temperature~\cite{lifshitz,kampen}. 
For a long time, the prevailing view was that Casimir/van der Waals (vdW) interactions between dielectrically similar bodies are always attractive~\cite{milonni,milton,bordag,parsegian}, and in fact constitute a major cause of stiction and non-contact friction in microsized devices~\cite{allen}. These issues have motivated the search for ways to make Casimir/vdW interactions less attractive or even repulsive. 
The search for repulsive Casimir/vdW forces is circumscribed by the existence of a no-go theorem, which states that the Casimir/vdW force in a mirror-symmetric setup is always attractive~\cite{kenneth-klich,bachas,rahi}. 
In the context of conventional dielectric media, one way to evade the theorem and generate a repulsive vdW force between two coplanar layers across a gap of a different material is to have the dielectric permittivities increase (or decrease) across the layers~\cite{parsegian,esteso1,esteso2}. However, such a scenario does not apply to systems involving a pair of dielectrics separated by a vacuum gap, which are the systems that would be relevant to device application~\cite{bordag}. 

Recently, it was recognized that two dielectrically similar layers can repel each other across a vacuum gap via Casimir/vdW forces, if these layers are made of topological insulator (TI) material~\cite{hasan-kane,qi1,qi2,qi3,qi4,ando,shen,bernevig,franz-molenkamp}. Besides being characterized electromagnetically by the dielectric permittivity and the magnetic permeability, TI materials are also characterized by a third parameter, that of the magnetoelectric polarizability or theta angle $\vartheta$, which can give rise to exotic electrodynamic behavior such as the induction of an image magnetic monopole by an electric charge near a TI surface~\cite{qi3}. By tuning this parameter, one can achieve ``Casimir repulsion", and physically such repulsion is a consequence of polarization mixing induced by the magnetoelectric coupling~\cite{grushin-cortijo}. Such Casimir/vdW repulsion was studied for dielectrically isotropic TIs~\cite{grushin-cortijo,nie,chen1} and TIs which are dielectrically anisotropic and whose optic axes are aligned with each other~\cite{grushin,pablo,chen2}.

The question thus arises as to the character of the vdW interaction between dielectrically anisotropic TI slabs whose optic axes are not aligned with each other. This introduces, firstly, the possibility of a vdW \emph{torque}~\cite{barash,parsegian-weiss,lu-podgornik,munday1,munday2}, and secondly, the possibility of tuning the character of the vdW \emph{force} via tuning the angular difference between the optic axes~\cite{lu-podgornik}. Our work sets out to address this two-fold question, in particular, examining the case where the TI slabs possess \emph{uniaxial anisotropy} (i.e., the dielectric permittivity along the optic axis direction is different from the one characterizing the two directions that are perpendicular to it, and the dielectric permittivities along the two remaining directions are identical), and the optic axes of both of the slabs are perpendicular to the direction normal to the planes of the slab surfaces which are facing each other (cf. Fig.~\ref{fig:slabs}). 

%For two ordinary dielectric slabs whose optical anisotropy is small and which are separated by an dielectrically isotropic gap, the vdW torque is proportional to the square of the optical anisotropy $\gamma$, is always restoring (i.e., tends to align the optic axes of the two slabs), and decays as the square of the separation distance, in the non-retardation regime. 

The problem of van der Waals torque was first studied in Refs.~\cite{kats} and \cite{barash0}, and a correct (albeit complex) calculation which accounts for retardation effects was later presented in Ref.~\cite{barash}. The complexity of the retarded solution arises because the crystal anisotropy gives rise to \emph{two} different types of waves, viz., ordinary waves and extraordinary waves, which are characterized by their different wavenumbers~\cite{barash}. On the other hand, for isotropic dielectrics~\cite{agranovich,LL8} and for anisotropic crystals in the non-retarded limit~\cite{parsegian-weiss}, there is only \emph{one} type of wave. 
A solution was obtained in Ref.~\cite{parsegian-weiss} for the latter case, which is simpler as it made use of the Maxwell equations in the non-retarded limit, and the solution was shown to be identical to the non-retarded limit of the solution obtained in Ref.~\cite{barash}. Building on these works, the character of the vdW torque was also investigated (in the non-retarded limit) for multilayered birefringent slab systems~\cite{lu-podgornik}, and proposals for the detection of the vdW torque have also been made, such as suspending a birefringent disk on a layer of ethanol above a barium titanate plate~\cite{munday1}, or measuring the rotation of a cholesteric liquid crystal induced by the vdW torque from a birefringent crystal~\cite{munday2}. One expects that a theoretical investigation of the vdW torque is also important for the application of rotating nanoelectromechanical systems~\cite{kim}, as this involves potential issues of frictional torque. If the vdW torque is small for TI rotors then one expects a corresponding smallness for the frictional torque. 

In this paper, we generalize the non-retardation approach of Parsegian and Weiss~\cite{parsegian-weiss} to the case of topological insulators. 
Our adoption of the non-retardation approach is motivated, firstly, by the complexity of the retarded calculation of the vdW torque, which is already evident for the case of ordinary dielectrics~\cite{barash}, and expected to be even more so for the case of TI slabs; secondly, by the fact that one can use the result from a non-retardation approach to check the result from a full, retarded calculation if the latter were carried out. In our work we derive the non-retarded vdW free energy for anisotropic TI slabs, which agrees with the result of Ref.~\cite{grushin} for the case of dielectric isotropy, thus confirming the essential consistency of both approaches. 
Owing to the magnetoelectric coupling present at each dielectric interface, a non-vanishing and spatially varying vector potential is induced even in the absence of currents, thereby enriching the physics. In terms of the force, we find that the mixing of the scalar and vector potentials in the problem leads to the possibility of vdW repulsion in the non-retarded or static limit, consistent with the findings of Ref.~\cite{grushin}. 
In Sec.~II, we present the Maxwell equations in the non-retarded limit. In Sec.~III, we derive the boundary conditions. In Sec.~IV, we compute the secular determinant, from which we derive the vdW free energy. We then compute and study the behavior of the vdW force and torque in Sec.~V. Finally, Sec.~VI contains a discussion of our results and our conclusions. 

\section{Field equations}

In the non-retardation regime (where $c \rightarrow \infty$), the modified Maxwell equations for an isotropic, homogeneous topological insulator (TI) are given by~\cite{wilczek,huerta1,huerta2,martin-ruiz1,martin-ruiz2} 
\begin{subequations}
\label{maxwell}
\ba
\nabla\!\cdot\! ({ \bf \varepsilon \, \bf E}) &=& \nabla \bar{\alpha} \cdot {\bf B} 
\\
\nabla\!\times\!{\bf B} &=& - \nabla \bar{\alpha} \times {\bf E} 
\\
\nabla\!\cdot\!{\bf B} &=& 0
\\
\nabla\!\times\!{\bf E} &=& 0
\ea
\end{subequations}
Here, $\varepsilon$ is the dielectric permittivity {\em tensor}, given by 
\ba
&&\varepsilon_i(\theta_i)
\\
&=\!& 
\begin{pmatrix}
\epsilon_{ix} \cos^2\theta_i + \epsilon_{iy} \sin^2\theta_i & (\epsilon_{ix} - \epsilon_{iy}) \sin\theta_i\cos\theta_i & 0
\\
(\epsilon_{ix} - \epsilon_{iy}) \sin\theta_i\cos\theta_i & \epsilon_{iy} \cos^2\theta_i + \epsilon_{ix} \sin^2\theta_i & 0
\\
0 & 0 & \epsilon_{iz}
\end{pmatrix}
\nonumber
\ea  
The index $i=1,2,3$ labels the slabs, and $\theta_i$ (not to be confused with the theta angle $\vartheta$ in the topological magnetoelectric coupling) denotes the orientation of the optic axis of slab $i$ relative to a reference axis. 
The symbols $\epsilon_{ix}$, $\epsilon_{iy}$ and $\epsilon_{iz}$ denote the values of the dielectric permittivities along the three principal directions of the TI. 
For dispersive materials, the dielectric permittivity is also a function of the frequency, $\epsilon = \epsilon(\omega)$. 
We set the optic axis of slab 1 as our reference axis, i.e., $\theta_1 = 0$. We take slab 2 to be a vacuum gap, which is dielectrically isotropic. We thus have $\epsilon_{2x}=\epsilon_{2y}=\epsilon_{2z} \equiv \epsilon_{2z}$, and we have that $\varepsilon_2 = \epsilon_{2z} \mathbb{I}$, where $\mathbb{I}$ is the identity matrix.  

The symbol $\bar{\alpha}$ denotes the topological magnetoelectric coupling strength~\cite{qi2,essin}, whose value is constant in the bulk regions, but which changes discontinuously across a dielectric interface. 
Magnetoelectric couplings can arise via various types of mechanisms and materials, such as the elastic interaction of ferroelectric and ferromagnetic domains in composite multiferroics~\cite{fiebig,spaldin}, and the magnetoelectric coupling generally also depends on the temperature~\cite{hehl1,hehl2} and frequency~\cite{brown,hornreich}. 
In TIs, owing to a quantum Hall effect at the boundary~\cite{qi3}, the magnetoelectric coupling is quantized for frequencies small compared to the surface gap, i.e., $\vartheta=(2n+1)\pi$ ($n \in \mathbb{Z}$)~\cite{vazifeh-franz}. On the other hand, for larger frequencies the quantization breaks down, and the behavior of the TI reverts to that of an ordinary dielectric~\cite{grushin-dejuan,pablo-grushin}. 

\begin{figure}[h]
\centering
  \includegraphics[width=0.27\textwidth]{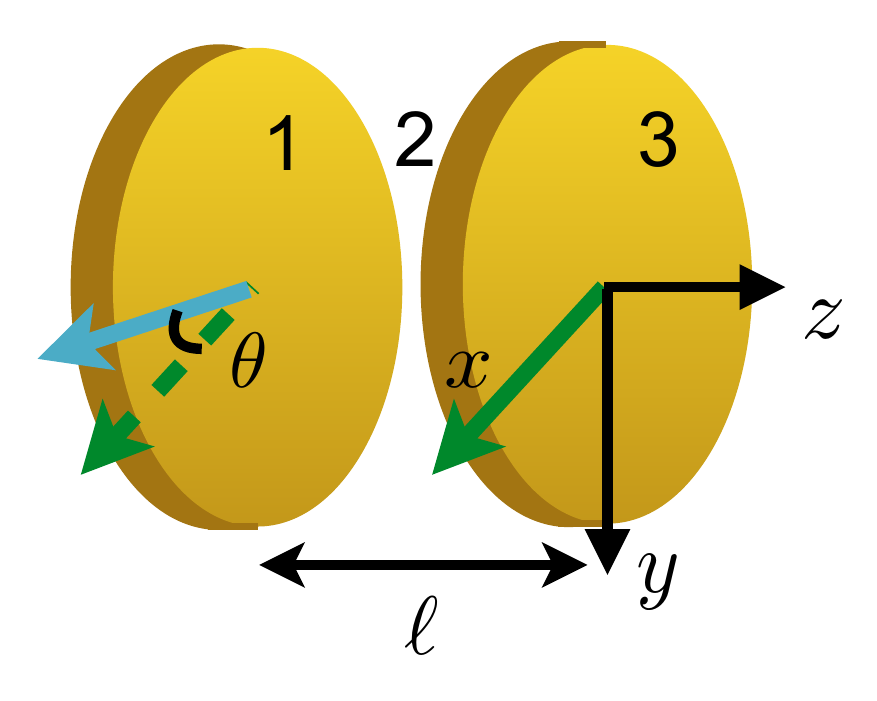}
  \caption{A pair of flat, coplanar topological insulator slabs (labeled 1 and 3) separated by a vacuum gap (labeled 2) of width $\ell$. The optic axis of slab 1 (3) is colored cyan (green), and the angular difference between the orientations of the optic axes is $\theta$. In the figure we have only shown a finite slice of the slabs, which are assumed to have infinitely large thicknesses and cross-sectional areas.} 
  \label{fig:slabs}
\end{figure}
For our TI-gap system (cf. Fig.~\ref{fig:slabs}), where slab 1 is defined as the region $z \leq 0$, slab 2 (the vacuum gap) as the region $0 < z < \ell$, and slab 3 as the region $z \geq \ell$, $\bar{\alpha}_2 = 0$ and $\bar{\alpha}_1$ and $\bar{\alpha}_3$ have non-zero values. Owing to the broken translational symmetry along the $z$-direction, we can see that $\bar{\alpha}$ has a spatial dependence only on $z$. 
Assuming that $\bar{\alpha}$ is quantized (we defer a discussion of the validity of this assumption to Sec.~IV), we can write $\bar{\alpha}_i = (2n+1) \, \alpha_i$ ($i=1,3$ are the labels for the TI slabs), and making use of the fact that the derivative of a Heaviside function is a Dirac delta-function, we have that 
\be
\partial_z \bar{\alpha}(z) = (2n+1) \,\alpha\, \delta(z) 
\ee
The last two modified Maxwell equations imply that we can express the ${\bf E}$ and ${\bf B}$ fields in terms of the scalar and vector potentials, viz., 
\begin{subequations}
\label{EA}
\ba
{\bf E} &=& - \nabla \phi,
\\
{\bf B} &=& \nabla \times {\bf A}
\ea
\end{subequations}
In the bulk regions, $\partial_z \bar{\alpha} = 0$, and the modified Maxwell equations simply become the Maxwell equations, which are known to be invariant under a gauge transformation. We are thus free to impose a gauge condition. For convenience, we impose the Coulomb gauge condition, i.e., $\nabla \cdot {\bf A} = 0$. 
By substituting Eqs.~(\ref{EA}) into the first two modified Maxwell equations, we obtain
\begin{subequations}
\ba
-\nabla\!\cdot\! (\varepsilon \nabla \phi) &=& \nabla \bar{\alpha} \cdot \nabla \times {\bf A}, 
\label{anne1}
\\
\nabla^2 {\bf A} &=& -\nabla \bar{\alpha} \times \nabla \phi
\label{anne2}
\ea
\end{subequations}
In the bulk regions, the equations for $\phi$ and ${\bf A}$ decouple and become homogeneous wave equations. To solve for $\phi$ in the bulk region, we postulate the following {\it Ansatz} (following Ref.~\cite{parsegian-weiss}): 
\be
\phi_i(\rv) = \!\int\!\! \frac{d^2\qv}{(2\pi)^2} \, f_i(z) \, {\rm e}^{i (q_x x + q_y y)}
\ee
Plugging this into Eq.~(\ref{anne1}) with $\bar{\alpha}=0$ yields
\be
\partial_z^2 f_i(z) - \rho_i^2(\theta) \, f_i(z) = 0,    
\label{d2f}
\ee
where 
\be
\rho_i^2(\theta_i) \equiv \frac{\epsilon_{ix}}{\epsilon_{iz}} (q_x \cos\theta_i+q_y\sin\theta_i)^2 
+ \frac{\epsilon_{iy}}{\epsilon_{iz}} (q_y \cos\theta_i-q_x\sin\theta_i)^2
\label{rho}
\ee
By making use of circular symmetry, we can express $q_x$ and $q_y$ in terms of a new angular variable $\psi$ (which is not to be confused with the optic axis orientation $\theta$), i.e., $q_x = q \cos \psi$, $q_y = q \sin \psi$, whence we also obtain a relation between the corresponding integration measures, $\int dq_x \, dq_y = \int d\psi\,dq\,q$~\cite{parsegian-weiss}. We can therefore express $\rho_i$ (Eq.~(\ref{rho})) in the manner~\cite{lu-podgornik}
\be
\rho_i(\theta_i) = q \, g_i(\theta_i-\psi), 
\ee
where $g_i(\theta-\psi)$ encodes the effects of dielectric anisotropy, and is given by 
\be
g_i(\theta-\psi) \equiv \sqrt{\frac{\epsilon_{iy}}{\epsilon_{iz}} + \frac{\epsilon_{ix} - \epsilon_{iy}}{\epsilon_{iz}} \cos^2 (\theta - \psi)} 
\ee 
We can easily check that for a dielectrically isotropic slab, $g_i \rightarrow 1$ and $\rho_i \rightarrow q \equiv (q_x^2+q_y^2)^{1/2}$.  
We shall consider the case where $\epsilon_{iy} = \epsilon_{iz}$, which enables the formula for $g_i$ to be further simplified. If we introduce the parameter 
\be
\gamma_{i} \equiv \frac{\epsilon_{ix}}{\epsilon_{iz}} - 1,
\ee
the anisotropy factor becomes 
\be
g_i = 
\sqrt{1 + \gamma_i \cos^2(\theta_i - \psi)}
\ee
For weak anisotropy (i.e., small dielectric difference between the optic axis and the ordinary axes), $\gamma_i \ll 1$, and we have 
\be
g_i 
\approx
1 + \frac{1}{2}\gamma_i \cos^2(\theta_i - \psi),
\ee
For slabs of the same dielectric material, $\gamma_1 = \gamma_3 \equiv \gamma$. For the vacuum gap ($i=2$), $\gamma_2 = 0$ and $g_2 = 1$.  

Turning back to Eq.~(\ref{d2f}), the solution takes the form 
\ba
f_1 (z) =& a_1 \, {\rm e}^{\,\rho_1 z} &(z \leq 0)
\nonumber\\
f_2 (z) =& a_2 \, {\rm e}^{\,\rho_2 z} + b_2 \, {\rm e}^{- \rho_2 z} &(0 < z < \ell)
\nonumber\\
f_3 (z) =& b_3 \, {\rm e}^{- \rho_3 z} &(z \geq \ell)
\label{f}
\ea
Similarly, as ${\bf A}$ satisfies a homogeneous wave equation in the bulk (cf. Eq.~(\ref{anne2})), we can postulate the {\it Ansatz}: 
\be
{\bf A}_i(\rv) = \!\int\!\! \frac{d^2\qv}{(2\pi)^2} \, \hv_i(z) \, {\rm e}^{i(q_x x + q_y y)}
\ee
Plugging this into Eq.~(\ref{anne2}) yields 
\ba
\hv_1 (z) =& \cv_1 \, {\rm e}^{\,q z} &(z \leq 0)
\nonumber\\
\hv_2 (z) =& \cv_2 \, {\rm e}^{\,q z} + \dv_2 \, {\rm e}^{- q z} &(0 < z < \ell)
\nonumber\\
\hv_3 (z) =& \dv_3 \, {\rm e}^{- q z} &(z \geq \ell)
\label{h}
\ea

\section{Boundary conditions}

Our next task is to derive the secular determinant from the boundary conditions. The boundary conditions for $\phi$ can be derived by integrating Eq.~(\ref{anne1}) over a narrow strip across each dielectric boundary, and requiring that $\phi$ is continuous across each dielectric boundary. We have 
\ba
&&\phi_1(0) = \phi_2(0),
\,\,\,  
\phi_2(\ell) = \phi_3(\ell),
\label{sp1}\\
&&[\epsilon_{1z} \partial_z \phi_1 - \epsilon_{2z} \partial_z \phi_2 ]_{z=0}
= -[\bar{\alpha}_1 (\partial_x A_{1y} - \partial_y A_{1x})]_{z=0},
\nonumber\\
&&[\epsilon_{2z} \partial_z \phi_2 - \epsilon_{3z} \partial_z \phi_3 ]_{z=\ell} 
= [\bar{\alpha}_3 (\partial_x A_{3y} - \partial_y A_{3x})]_{z=\ell}
\nonumber
\ea
We can derive the boundary conditions for $\Av$ by an analogous method, via integrating Eq.~(\ref{anne2}) and requiring the continuity of $\Av$ across each dielectric boundary. In component form, Eq.~(\ref{anne2}) become
\ba
\nabla^2 A_x &=& \bar{\alpha} \, \delta(z-z_m) \, \partial_y \phi,
\nonumber\\
\nabla^2 A_y &=& - \bar{\alpha} \, \delta(z-z_m) \, \partial_x \phi,
\nonumber\\
\nabla^2 A_z &=& 0
\ea
where $z_m$ ($m=L, R$) denote the position of the dielectric interfaces, i.e., $z_L=0$ and $z_R=\ell$. 
Integrating across the dielectric boundary at $z=0$ leads to 
\ba
\left[ \partial_z A_{2x} - \partial_z A_{1x} \right]_{z=0} 
&=& -\bar{\alpha}_1 \, \partial_y \phi_1(z=0),
\nonumber\\
\left[ \partial_z A_{2y} - \partial_z A_{1y} \right]_{z=0} 
&=& \bar{\alpha}_1 \, \partial_x \phi_1(z=0),
\nonumber\\
\left[ \partial_z A_{2z} - \partial_z A_{1z} \right]_{z=0}  &=& 0
\label{vp1}
\ea
Integrating across $z=\ell$ leads to
\ba
\left[ \partial_z A_{3x} - \partial_z A_{2x} \right]_{z=\ell} 
&=& \bar{\alpha}_3 \, \partial_y \phi_3(z=\ell),
\nonumber\\
\left[ \partial_z A_{3y} - \partial_z A_{2y} \right]_{z=\ell} 
&=& -\bar{\alpha}_3 \, \partial_x \phi_3(z=\ell),
\nonumber\\
\left[ \partial_z A_{3z} - \partial_z A_{2z} \right]_{z=\ell}  &=& 0
\label{vp2}
\ea
The above six equations are supplemented by six additional equations the describe the continuity of $\Av$ across each dielectric boundary, viz., 
\begin{subequations}
\label{vp3}
\ba
\Av_1(z=0) &=& \Av_2(z=0),
\\
\Av_2(z=\ell) &=& \Av_3(z=\ell)
\ea
\end{subequations} 
Equations~(\ref{sp1}), (\ref{vp1}), (\ref{vp2}) and (\ref{vp3}) can be expressed in terms of the coefficients $a_1$, $a_2$, $b_2$, $b_3$, $\cv_1$, $\cv_2$, $\dv_2$ and $\dv_3$. Using Eqs.~(\ref{f}) and (\ref{h}), we can rewrite Eqs.~(\ref{sp1}) as 
\ba
&&a_1 = a_2 + b_2,
\nonumber\\
&&b_3 \, {\rm e}^{-\rho_3 \ell}
= a_2 \, {\rm e}^{\rho_2 \ell} + b_2 \, {\rm e}^{-\rho_2 \ell},
\nonumber\\
&&\epsilon_{1z} \rho_1 a_1 - \epsilon_{2z} \rho_2 (a_2 - b_2) 
= -i \, \bar{\alpha}_1 (q_x c_{1y} - q_y c_{1x}),
\nonumber\\
&&\epsilon_{2z} \rho_2 (a_2 \, {\rm e}^{\rho_2 \ell} - b_2 \, {\rm e}^{- \rho_2 \ell}) + \epsilon_{3z} \rho_3 b_3 \, {\rm e}^{-\rho_3 \ell} 
\nonumber\\
&&\quad= i\,\bar{\alpha}_3 (q_x d_{3y} - q_y d_{3x}) \, {\rm e}^{-q \ell}
\label{sp1'}
\ea
Equations~(\ref{vp1}) become
\ba
&&c_{2x} - d_{2x} - c_{1x} = -i\,(\bar{\alpha}_1/q)\,q_y a_1,
\nonumber\\
&&c_{2y} - d_{2y} - c_{1y} = i\,(\bar{\alpha}_1/q)\,q_x a_1,
\nonumber\\
&&c_{2z} - d_{2z} - c_{1z} = 0
\label{vp1'}
\ea
Equations~(\ref{vp2}) become
\ba
&&-d_{3x} \, {\rm e}^{-q \ell} - c_{2x} \, {\rm e}^{\, q \ell} + d_{2x} \, {\rm e}^{-q \ell} 
= i\,(\bar{\alpha}_3/q)\,q_y b_3 \, {\rm e}^{-\rho_3 \ell}, 
\nonumber\\
&&-d_{3y} \, {\rm e}^{-q \ell} - c_{2y} \, {\rm e}^{\, q \ell} + d_{2y} \, {\rm e}^{-q \ell} 
= - i \,(\bar{\alpha}_3/q)\,q_x b_3 \, {\rm e}^{-\rho_3 \ell}, 
\nonumber\\
&&-d_{3z} \, {\rm e}^{-q \ell} - c_{2z} \, {\rm e}^{\, q \ell} + d_{2z} \, {\rm e}^{-q \ell} 
= 0
\label{vp2'}
\ea
Finally, Eqs.~(\ref{vp3}) become
\ba
&&\cv_1 = \cv_2 + \dv_2,  
\nonumber\\
&&
\cv_2 \, {\rm e}^{\, q \ell} + \dv_2 \, {\rm e}^{-q \ell} = \dv_3 \, {\rm e}^{-q \ell}
\label{vp3'}
\ea
We have sixteen unknown coefficients, viz., $a_1$, $a_2$, $b_2$, $b_3$, $\cv_1$, $\cv_2$, $\dv_2$ and $\dv_3$, and sixteen equations that describe the boundary conditions, viz., Eqs.~(\ref{sp1'}), (\ref{vp1'}), (\ref{vp2'}) and (\ref{vp3'}). The coefficients can thus be solved. For the purpose of deriving the vdW free energy, however, we do not need to know the explicit values of the coefficients, only the secular determinant of the equations. 

\section{Free energy}

Equations~(\ref{sp1'}) to (\ref{vp3'}) can be re-expressed in matrix form ${\bf M} \cdot {\bf r} = 0$, where ${{\bf r}}$ is a 16-component vector formed from the unknown amplitudes in the \emph{Ans\"{a}tze} Eqs.~(\ref{f}) and (\ref{h}), viz., 
\be
{{\bf r}}^{{\rm T}} = (a_1,a_2,b_2,b_3, {{\bf c_1}}^{{\rm T}}, {{\bf c_2}}^{{\rm T}}, {{\bf d_2}}^{{\rm T}}, {{\bf d_3}}^{{\rm T}}). 
\ee
The matrix $\bf{M}$ contains the coefficients of the amplitudes. 
Equations~(\ref{sp1'}) to (\ref{vp3'}) have a non-trivial solution if the determinant of the secular equation vanishes, i.e., $|{\bf M}| = 0$.  
A straightforward calculation shows that the determinant is proportional to $\mathcal{D}(\omega, q\ell)$, where 
\ba
\mathcal{D}(\omega, \xi) 
&\!\equiv\!& 
1 - \frac{8 \, \bar{\alpha}_1 \, \bar{\alpha}_3 \, \epsilon_{2z} g_2}{v_1 v_3} {\rm e}^{-(1+g_2) \xi} 
\label{D}
\\
&&- \frac{u_1 u_3}{v_{1} v_{3}} {\rm e}^{-2 g_2 \xi}
- \frac{ \bar{\alpha}_1^2 \, \bar{\alpha}_3^2}{v_1 v_3} 
\left( {\rm e}^{-2 \xi} - {\rm e}^{-2(1+g_2) \xi} \right)
\nonumber
\ea
The form of $\mathcal{D}$ is such that $\ln \mathcal{D} \rightarrow 0$ as $\xi \rightarrow \infty$. 
Here, $\xi$ is a free, dimensionless variable, not to be confused with the Matsubara frequency (which we will introduce shortly), and we have defined ($i=1,3$)
\begin{subequations}
\ba
u_i &\equiv& \bar{\alpha}_i^2 + 2(\epsilon_{iz} g_i(\theta_i) - \epsilon_{2z} g_2(\theta_2)), 
\label{ui}
\\
v_i &\equiv& \bar{\alpha}_i^2 + 2(\epsilon_{iz} g_i(\theta_i) + \epsilon_{2z} g_2(\theta_2)) 
\label{vi}
\ea
\end{subequations}
For the solution to be non-trivial, we require that the secular determinant vanishes, i.e., $\mathcal{D}(\omega,q\ell) = 0$. By the Argument Principle~\cite{barash-ginzburg,ginzburg}, the sum over the allowed frequencies of the electromagnetic fluctuations between the slabs in the free energy (which gives rise to the van der Waals interaction) can be expressed in the form of a contour integral over another function whose poles occur at the values of the allowed frequencies. This leads to the following form of the vdW interaction free energy~\cite{parsegian-weiss} 
\be
G(\ell) = k_{{\rm B}}T \, {\sum_{n=0}^{\infty}}' 
\!\int_0^{2\pi} \!\! \frac{d\psi}{2\pi} 
\!\int_0^\infty \!\! \frac{dq}{2\pi} \, q 
\ln \mathcal{D}(i\xi_n, q\ell),   
\label{G}
\ee
where $\xi_n$ are the Matsubara frequencies, defined by $\xi_n \equiv 2\pi n k_{{\rm B}}T/\hbar$, and the prime on the sum over Matsubara frequencies tells one that the $n=0$ term should be multiplied by an additional factor of $1/2$. 
By setting $\epsilon_{1z}=\epsilon_{3z}$, $\epsilon_{2z}=1$ and $\rho_1=\rho_2=\rho_3=q$ (which would be valid for a pair of isotropic TI slabs of the same material interacting across a vacuum gap), and taking $c \rightarrow \infty$, we recover the non-retarded limit of the vdW free energy result of Ref.~\cite{grushin} (see App.~A for details). 
Equation~(\ref{G}), supplemented by Eqs.~(\ref{D}), (\ref{ui}) and (\ref{vi}) constitute the central results of our paper. 

In practice, the non-retarded approximation works well if the separation distance smaller than all the resonant absorption wavelengths of the slab material~\cite{parsegian}. If we consider the TI material $\rm{TlBiSe}_2$, there are two resonances, viz., a phonon resonance near $56\,\rm{cm}^{-1}$ (which corresponds to an absorption wavelength of $180 \, \rm{\mu m}$), as well as a plasma resonance near $800\,\rm{cm}^{-1}$ (which corresponds to an absorption wavelength of $12.5 \, \rm{\mu m}$)~\cite{mitsas}. Thus, our non-retarded approximation should work well for $d \lesssim 12.5 \, \rm{\mu m}$. 

As we noted earlier, like the dielectric permittivity and magnetic permeability~\cite{geyer} the magnetoelectric coupling $\bar{\alpha}$ also depends on the frequency. It is known that $\bar{\alpha}$ is only quantized for low frequencies, and at larger frequencies the quantization is lost~\cite{grushin-dejuan,pablo-grushin}. The finite-frequency behavior of the magnetoelectric response has been studied for three-dimensional TIs~\cite{grushin-dejuan} and Chern insulators~\cite{pablo-grushin} at zero temperature, and found to vanish as the frequency becomes infinitely large. On the other hand, as far as we are aware of, the corresponding finite-temperature problem has not been studied. However, we expect that the magnetoelectric response should also become transparent for infinitely large frequencies. Thus, for $\xi_n < m/\hbar$ (where $m$ is the surface mass gap), we expect $\bar{\alpha}$ to be quantized, whereas for $\xi_n \gg m/\hbar$, we expect that $\bar{\alpha}$ becomes negligible. 

The above considerations suggest the scope in which we can apply our non-retarded result. Specifically, within the non-retarded regime, we can study the behavior of sufficiently hot systems, in which we can approximate the vdW free interaction by the zero-frequency Matsubara term (for which $\bar{\alpha}$ is quantized). The finite-frequency Matsubara terms, for which $\bar{\alpha}$ is not quantized, do not contribute significantly and can then be neglected. To obtain an estimate for the temperature above which the finite-frequency terms can be neglected, we estimate the Matsubara frequency above which the dielectric response and magnetoelectric response become near-transparent. 
For the dielectric response, we require $2\pi k_{{\rm B}}T/\hbar \gg \omega_{{\rm R}}$ (where $\omega_{{\rm R}}$ is the resonant absorption frequency), which gives $T \gg  \hbar\omega_{{\rm R}}/(2\pi k_{{\rm B}})$. 
For the magnetoelectric response, we require $2\pi k_{{\rm B}}T/\hbar \gg m/\hbar$, which gives
 $T \gg m/(2\pi k_{{\rm B}})$. To the best of our knowledge, we have not been able to find simultaneously values of $\omega_{{\rm R}}$ and $m$ for topological insulator materials. However, individually we have found, e.g., that $\omega_{{\rm R}} \approx 1.5 \times 10^{14} \, {{\rm s}}^{-1}$ for ${{\rm TlBiSe}}_2$ (using the plasma resonance frequency~\cite{mitsas} rather than the phonon resonance frequency, as we want the high-temperature limit to hold for both the phonon and plasmon resonances), and $m \approx 50 \, {{\rm meV}}$ for $\rm{Bi}_2\rm{Se}_3$~\cite{grushin-dejuan,chen,wray}. This implies that $T \gg 183 \,{{\rm K}}$ for the former and $T \gg 92 \,{{\rm K}}$ for the latter. For these temperature regimes, we can approximate the vdW interaction by its behavior in the $T \rightarrow \infty$ limit. 

\section{van der Waals force and torque}

\begin{figure*}[t]
\begin{center}
\begin{minipage}[b]{0.47\textwidth}
\includegraphics[width=\textwidth]{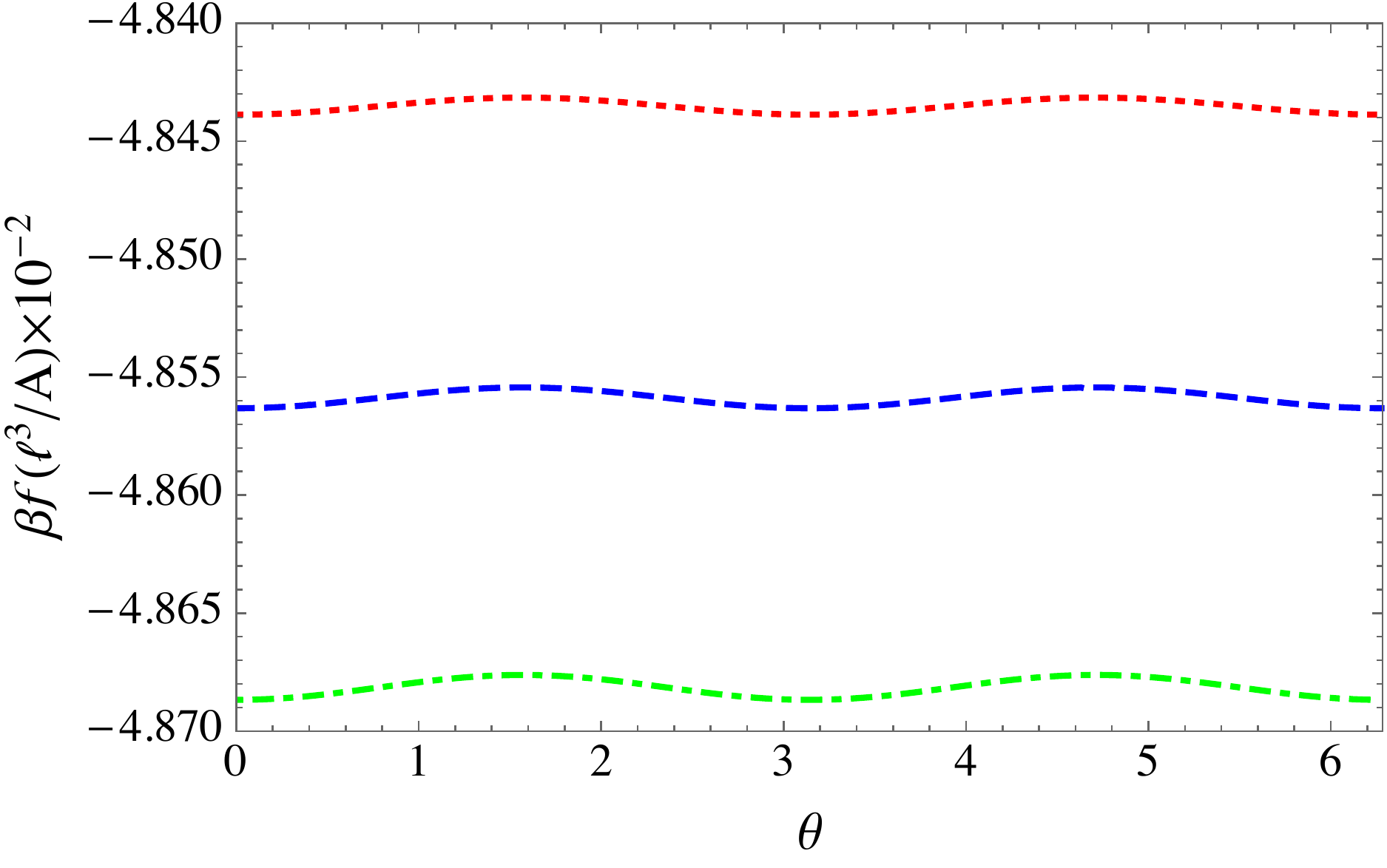} (a)
\end{minipage} 
\begin{minipage}[b]{0.47\textwidth}
\includegraphics[width=\textwidth]{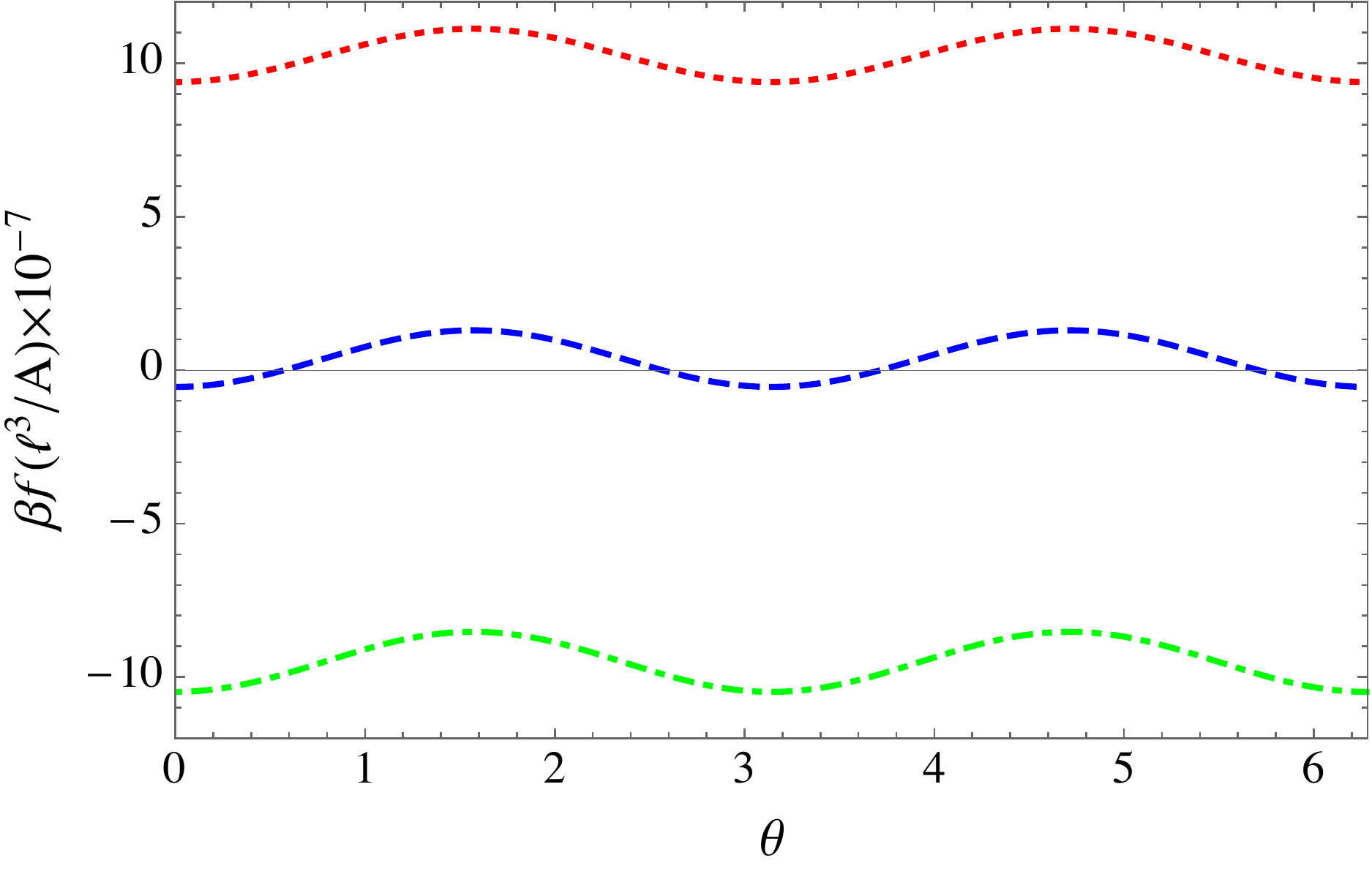} (b)
\end{minipage}
\end{center}

\caption{Behavior of the van der Waals force (in units of $A/\beta \ell^3$, where $A$ is the cross-sectional area of the slabs, $\beta$ is the inverse temperature and $\ell$ is the inter-slab separation) as a function of optic axis misalignment $\theta$, between a pair of dielectrically similar topological insulator slabs with the same dielectric permittivity, but different magnetoelectric polarizabilities $\bar{\alpha}$: (a)~$\bar{\alpha}_1=-\bar{\alpha}_3=3/137$ and static dielectric permittivity $\epsilon_{1z}(0)=\epsilon_{1y}(0)=\epsilon_{3z}(0)=\epsilon_{3y}(0)=4$; (b)~$\bar{\alpha}_1=-\bar{\alpha}_3=21/137$ and static dielectric permittivity $\epsilon_{1z}(0)=\epsilon_{1y}(0)=\epsilon_{3z}(0)=\epsilon_{3y}(0)=1.20179$. For both cases the intervening gap is a vacuum ($\epsilon_2=1$). In (a) the behavior is plotted for $\gamma=0.11$ (green, dot-dashed), $0.1$ (blue, dashed) and $0.09$ (red, dotted). In (b) the behavior is plotted for $\gamma=0.0103$ (green, dot-dashed), $0.01$ (blue, dashed) and $0.0097$ (red, dotted).}
 \label{fig:force}
 \end{figure*}
 
 \begin{figure}[h]
\centering
  \includegraphics[width=0.47\textwidth]{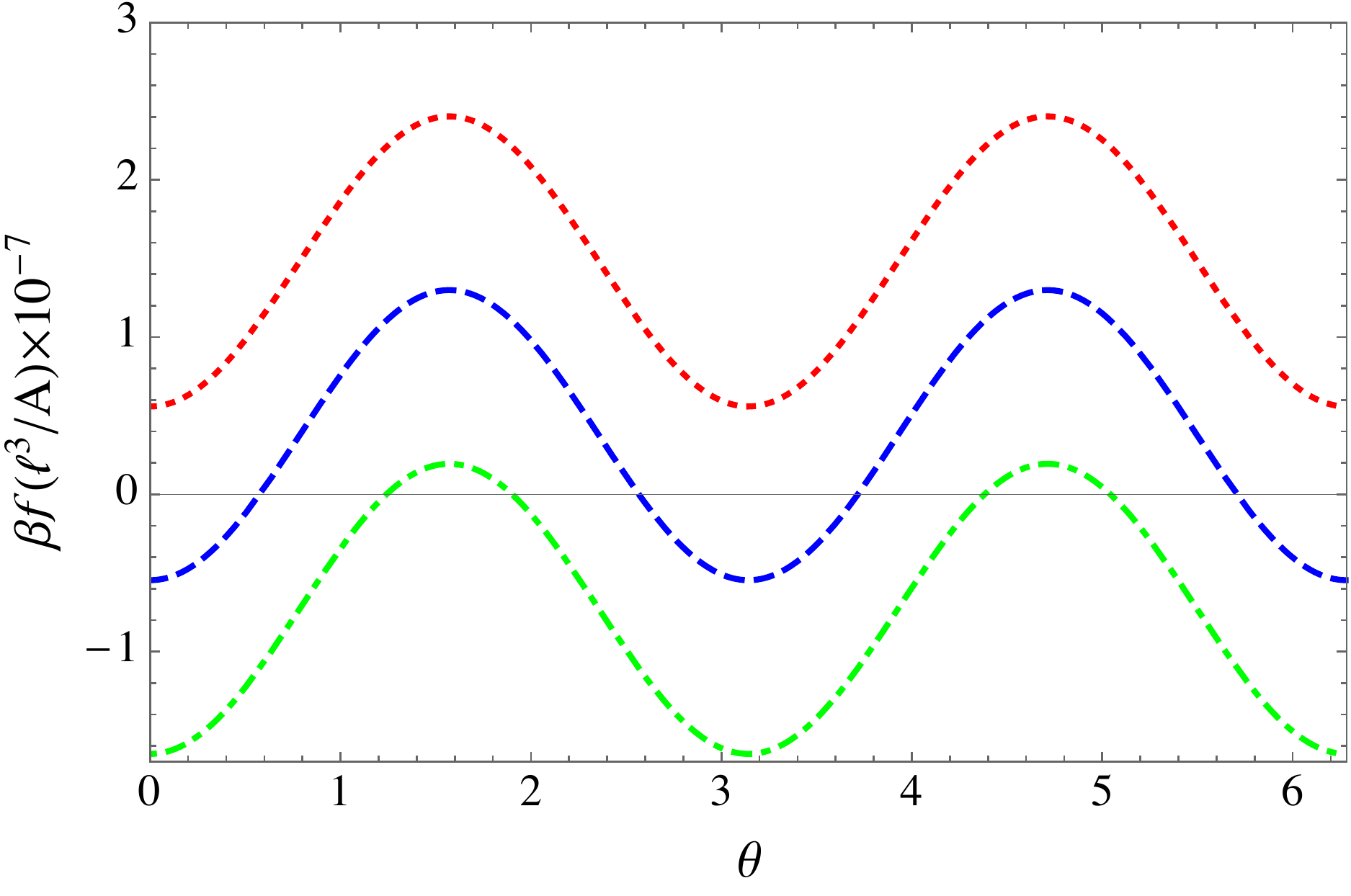}
  \caption{Behavior of the van der Waals force (in units of $A/\beta \ell^3$) as a function of optic axis misalignment $\theta$, for a pair of dielectrically similar topological insulator slabs with $\bar{\alpha}_1=-\bar{\alpha}_3=21/137$ and $\gamma = 0.01$, separated by a vacuum gap ($\epsilon_2 = 1$); the curves are plotted for static dielectric permittivity $\epsilon_{1z}(0)=\epsilon_{1y}(0)=\epsilon_{3z}(0)=\epsilon_{3y}(0)$ equal to the value of (i)~$1.2018$ (green, dot-dashed), (ii)~$1.20179$ (blue, dashed), and (iii)~$1.20178$ (red, dotted).}
  \label{fig:69perm}
\end{figure}
 
\subsection{Force}

From the free energy expression Eq.~(\ref{G}), we can derive the vdW force per unit area $f$ from the formula $f = -\partial G/\partial \ell$, which gives 
\ba
f(\theta, \ell) = \frac{k_{{\rm B}}T}{2 \pi \ell^3} 
{\sum_{n=0}^{\infty}}' \!\int_0^\infty\!\!\! 
d\xi \, \xi \ln \mathcal{D} ( i\xi_n, \xi)
\ea
In deriving this result we have rescaled the momentum variable so that the distance dependence comes out into the prefactor, which scales inversely as the cube of the distance. 
The scaling behavior is thus the same as that for ordinary dielectrics and isotropic TIs. 

As discussed in the previous section, we study the character of the vdW force between TI slabs interacting across a vacuum gap ($g_2=1, \epsilon_{2z}=1$) in the limit $T\rightarrow\infty$. This means that we can approximate the vdW interaction by the zero Matsubara frequency contribution, and replace $\epsilon$, $\gamma$ and $\bar{\alpha}$ by their static values. 
The force behavior as a function of the misalignment angle $\theta$ between the optic axes is plotted in Figs.~\ref{fig:force} and \ref{fig:69perm}. 
First, let us consider how the behavior of the vdW force changes as the anisotropy $\gamma$ is varied.  Figure~\ref{fig:force}a shows the behavior for two TI slabs with static dielectric permittivity of 4 (which would be close to the dielectric constant of $\rm{TlBiSe}_2$~\cite{grushin-cortijo}) and static magnetoelectric coupling $\bar{\alpha}_1 = -\bar{\alpha}_3 = 3/137$, whilst Fig.~\ref{fig:force}b shows the corresponding behavior for static dielectric permittivity of 1.20179 and static magnetoelectric coupling $\bar{\alpha}_1 = -\bar{\alpha}_3 = 21/137$. The choice of the signs is motivated by the finding in Ref.~\cite{grushin} that vdW repulsion is possible if TI slabs possess magnetoelectric couplings of the same magnitude but opposite signs, and the integer factor in $\bar{\alpha}$ counts the number of surface states in the TI, the typical number of surface states being of the order of 1. Thus Fig.~2a would describe the behavior of a more realistic system. For both cases we see that increasing the anisotropy has the effect of making the vdW force more attractive/less repulsive. In the case of Fig.~2b, we see that the vdW force can become repulsive for sufficiently weak anisotropy. This is consistent with the results of Ref.~\cite{grushin}, there it was found that for \emph{isotropic} TIs in the high-temperature limit, vdW repulsion becomes possible only for static dielectric permittivities smaller than a threshold value which is approximately 2. As increasing the anisotropy has the effect of making the vdW force more attractive, we should not expect to see vdW repulsion for any value of the anisotropy $\gamma$ if the static dielectric permittivity is larger than 2. On the other hand, Casimir/vdW repulsion should become a possibility for all finite values of the dielectric permittivity at low or zero temperature, on the basis of results of Ref.~\cite{grushin-cortijo} and \cite{grushin}. However, the zero-temperature limit falls outside the scope of our study, as it requires retardation effects to be taken into account. 

Figure~3 shows that the vdW force becomes less attractive/more repulsive as the static dielectric permittivity is decreased. From Figs.~2 and 3, with $\gamma \neq 0$, we see that the force oscillates with $\theta$, becoming most repulsive (or least attractive) for $\theta = (n+1/2) \pi$ (where $n \in \mathbb{Z}$) and least repulsive (or most attractive) for $\theta = n \pi$. 
Intriguingly, for certain ranges of values of $\gamma$ and $\bar{\alpha}$ we can also ``tune" the force to be attractive when $\theta$ is an integer multiple of $\pi$, and repulsive when $\theta$ is a half-integer multiple of $\pi$ (cf. the blue-colored curves in Fig.~\ref{fig:force}(b) and 3). The dependence of the vdW force on the orientation of the optic axis of the TI slab thus opens up a means to control the strength of stiction or repulsion between the components of a nanodevice. 

To summarise, in the high-temperature limit, the vdW force can become repulsive for sufficiently small values of $\gamma$ and/or sufficiently large values of $\bar{\alpha}$, if the static dielectric permittivity is smaller than a threshold value (which is around 2). 

\subsection{Torque}

\begin{figure}[h]
\centering
  \includegraphics[width=0.47\textwidth]{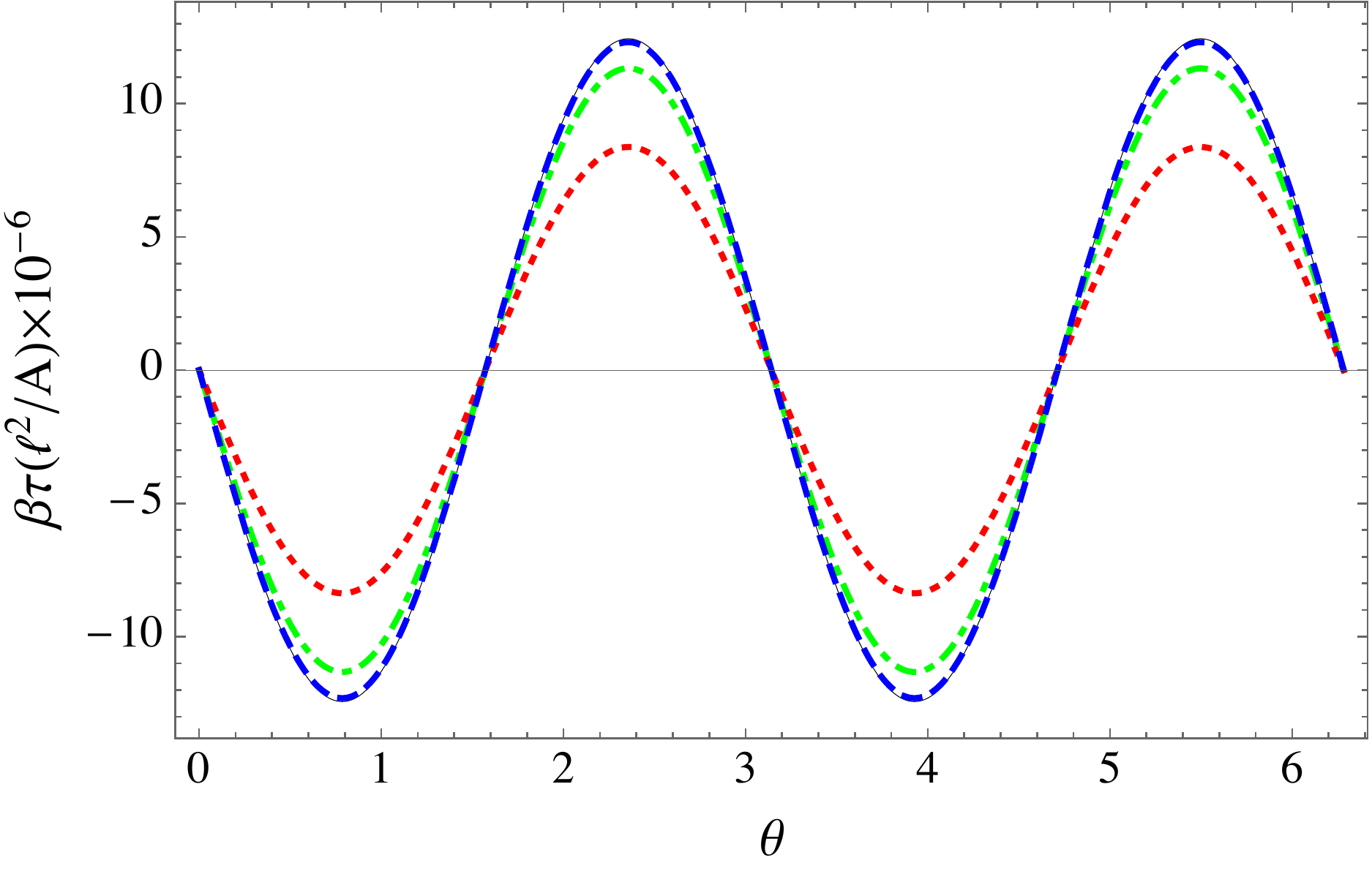}
  \caption{Behavior of the torque (multiplied by $\ell^2/A$, where $A$ is the cross-sectional area of each slab and $\ell$ is the inter-slab separation distance) for (i)~ordinary dielectric slabs (black, thin), (ii)~topological insulator slabs with $\bar{\alpha}_1 = \bar{\alpha}_3= 69/137$ (green, dashed), (iii)~topological insulator slabs with $\bar{\alpha}_1 = -\bar{\alpha}_3= 69/137$ (red, dot-dashed), and (iv)~topological insulator slabs with $\bar{\alpha}_1 = -\bar{\alpha}_3= 11/137$ (blue, dashed). We take the slabs to be dielectrically uniaxial, with static dielectric permittivity $\epsilon_z(0)=\epsilon_y(0)=4$ and $\gamma=0.2$, and the gap to be a vacuum ($\epsilon_2=1$).} 
  \label{fig:torque}
\end{figure}

Having explored the behavior of the vdW force, we turn next to the behavior of the vdW torque experienced by the TI slabs, which is given by
\be
\tau(\theta,\ell) = -\frac{\partial G}{\partial \theta}
\label{tau}
\ee
Let us consider a similar type of system as the one just analyzed for the force, where two TI slabs of the same material are separated by a vacuum gap. 
As in our analysis of the vdW force, we specialize to the limit $T\rightarrow\infty$. 
To make our calculations analytically tractable, let us expand $\ln \mathcal{D}$ in Eq.~(\ref{G}) in powers of $\gamma$ and ${\rm e}^{- 2 q \ell}$ to quadratic order, and perform the integrations $\int d\psi$ and $\int q \, d\psi$. To find the torque we apply Eq.~(\ref{tau}). We obtain the following result:
\ba
\label{tau1}
\tau(\theta,\ell) &\!\!\approx\!\!& 
-\frac{\gamma^2 (1+\frac{1}{4}\bar{\alpha}_1 \bar{\alpha}_3) \sin 2\theta}{1024\pi (1+\frac{1}{4}\bar{\alpha}_1^2)^3 (1+\frac{1}{4}\bar{\alpha}_3^2)^3 \ell^2} 
\nonumber\\
&&\times 
\Big( 
1 + \frac{1}{4}(\bar{\alpha}_1^2 + \bar{\alpha}_3^2) + \frac{3}{8}\bar{\alpha}_1\bar{\alpha}_3 
+ \frac{3}{16}\bar{\alpha}_1^2\bar{\alpha}_3^2 
\nonumber\\
&&\qquad
+ \frac{1}{32}\bar{\alpha}_1\bar{\alpha}_3(\bar{\alpha}_1^2+\bar{\alpha}_3^2) 
+\frac{3}{128}\bar{\alpha}_1^3\bar{\alpha}_3^3
\Big)
\ea
In the above, the parameters $\bar{\alpha}$ and $\gamma$ refer to their static values, i.e., $\bar{\alpha}(0)$ and $\gamma(0)$. 
We can check that in the limit of zero magnetoelectric coupling, we recover the known result of Ref.~\cite{parsegian-weiss}. 
We plot the behavior in Fig.~\ref{fig:torque} for TI slabs with $|\bar{\alpha}|$ equal to $69/137$ and $11/137$. We find that the vdW torque for a pair of TI slabs has the same sign as that for a pair of ordinary dielectrics, the sign being such that the torque is restoring when $\theta$ is a small perturbation from an integer multiple of $\pi$, and half-integer values of $\pi$ represent unstable configurations. For TI slabs the strength of the torque is generally weaker than that for ordinary dielectrics, which we can interpret as a consequence of the magnetoelectric coupling. For weaker magnetoelectric couplings (e.g., the system represented by the thin black curve in Fig.~4), the torque behavior virtually coincides with that of ordinary dielectrics, and differences in torque behavior between TIs and ordinary dielectrics become noticeable only for larger magnetoelectric couplings. 
The torque for TI slabs with $\bar{\alpha}$ of the same magnitude but opposite signs is weaker than that for TI slabs with $\bar{\alpha}$ of the same magnitude and sign. The enhanced weakening of the torque for the case where $\bar{\alpha}_1 = -\bar{\alpha}_3$ compared to the case where $\bar{\alpha}_1 = \bar{\alpha}_3$ appears to be associated with the emergence of a repulsive vdW force in the former case (compared to an attractive vdW force which arises in the latter case). Physically, if the forces were attractive, the slabs would experience a stronger moment of force that seeks to restore them to the state of mechanical equilibrium. 
From Eq.~(\ref{tau1}), we can also deduce that an increase in the strength of the magnetoelectric coupling $\bar{\alpha}$ has the effect of decreasing the magnitude of the vdW torque. 

\section{Discussion and conclusion}

In this paper, we have investigated the high-temperature limit of the van der Waals interaction between a pair of flat, coplanar topological insulator slabs with uniaxial dielectric anisotropy and separated by a vacuum gap, in particular looking into the character of the vdW force and torque, and how they vary as functions of the magnetoelectric coupling strength and the misalignment angle between the optic axes of the slabs. 
In addition to confirming that the vdW force can become repulsive for sufficiently large magnetoelectric coupling, we have also found that anisotropy can influence the sign of the vdW force. 
A repulsive vdW force can become attractive if the anisotropy is increased sufficiently. 
Furthermore, we have found that the force oscillates as a function of the misalignment angle between the optic axes of the TI slabs, being most repulsive/least attractive (least repulsive/most attractive) for angular differences that are integer (half-integer) multiples of $\pi$. For certain values of the permittivity and magnetoelectric coupling, it is also possible to tune the vdW force from attractive to repulsive as one varies the angular difference between the optic axes.
We have also found that the vdW torque is generally weaker for the case of TI slabs than for the case of ordinary dielectric ones. 
For the purpose of experimentally detecting and measuring the vdW torque, this would indicate that using ordinary dielectric micro-disks is more suitable than TI micro-disks~\cite{munday1,munday2}. 

In the high-temperature regime, it is known that the dielectric permittivities of the TI slabs have to be smaller than a threshold value (of approximately 2) to make vdW repulsion possible, whilst repulsion can occur for larger values of the dielectric permittivity in the low-temperature regime~\cite{grushin}. Current topological materials appear to possess static dielectric permittivities larger than 2. As permittivity depends on dipole density, it could be possible to make the permittivities smaller by increasing the porosity of the materials. 
Our calculation in the high-temperature limit also functions as a proof of principle calculation. In order to obtain more realistic predictions, it would have to be amended by taking into account the full frequency dependence of both the dielectric function and the magnetoelectric response. 

To address the low-temperature behavior, one can no longer neglect the finite-frequency Matsubara terms, and the magnetoelectric response at finite frequencies would have to be included. This requires a finite-temperature generalization of the magnetoelectric response derived in Refs.~\cite{grushin-dejuan,pablo-grushin}, which was calculated for the case of zero temperature. To probe the behavior at larger distances and/or the zero-temperature limit, one would also need to account for retardation effects. Retardation effects can be studied, for example, along the lines of Ref.~\cite{barash}. However, we expect the calculation to be more complicated, as one would have to consider contributions from both ordinary and extraordinary waves which emerge from the crystal anisotropy~\cite{agranovich,LL8}.  

\section{Acknowledgments}

The author thanks Rudolf Podgornik (formerly at the Institut Jo\v{z}ef Stefan, Ljubljana, and currently at the Institute of Physics, Chinese Academy of Sciences, Beijing) and David S. Dean (Universit\'{e} de Bordeaux) for showing him the garden of Casimir/van der Waals physics. He also thanks the three anonymous referees for their constructive comments. 

\appendix 

\section{Proof of the equivalence with the non-retarded limit of the result of Ref.~\cite{grushin}}

In Ref.~\cite{grushin}, the non-retarded limit of the vdW free energy for isotropic TIs ($\epsilon_\perp = \epsilon_z \equiv \epsilon$) can be deduced from Eqs.~(12) and (21) of the paper. In our notation, their result takes the form 
\be
\frac{G_g(\ell)}{A} = k_{{\rm B}}T \, 
{\sum_{n=0}^{\infty}}' \!\int\! \frac{d^2\qv}{(2\pi)^2} 
\ln \det \big( \mathbb{I} - \mathbb{R}_1 \cdot \mathbb{R}_3 \, {\rm e}^{-2q\ell} \big), 
\label{grushin}
\ee
where the reflection coefficient matrices are defined by ($i=1,3$)
\be
\mathbb{R}_i = \frac{1}{\bar{\alpha}_i^2 + 2\epsilon + 2} 
\begin{pmatrix}
-\bar{\alpha}_i^2 & 2\bar{\alpha}_i \\
2\bar{\alpha}_i & \bar{\alpha}_i^2 + 2\epsilon - 2
\end{pmatrix}
\ee
Evaluating the determinant, we obtain 
\ba
&&\frac{G_g(\ell)}{A} = 
k_{{\rm B}}T \, 
{\sum_{n=0}^{\infty}}' 
\!\int\! \frac{d^2\qv}{(2\pi)^2} 
\nonumber\\
&&\quad\times\ln \bigg(  
1 - \frac{u_1 u_3 + 8 \bar{\alpha}_1 \bar{\alpha}_3 + \bar{\alpha}_1^2 \bar{\alpha}_3^2}{v_1 v_3} \, {\rm e}^{-2q\ell} 
\nonumber\\
&&\qquad
+ \frac{\bar{\alpha}_1^2 \bar{\alpha}_3^2}{v_1^2 v_3^2} \big( u_1 u_3 + 4(u_1 + u_3) + 16 \big) \, {\rm e}^{-4q\ell}
\bigg) 
\label{a3}
\ea
Here, $\epsilon$ is the value of the dielectric permittivity of each slab, and $u_i = \bar{\alpha}_i^2 + 2\epsilon - 2$ and $v_i = \bar{\alpha}_i^2 + 2\epsilon + 2$ for a pair of isotropic TI slabs separated by a vacuum gap. It is easy to prove the algebraic identity
\be
v_1 v_3 = u_1 u_3 + 4(u_1 + u_3) + 16
\ee
Equation~(\ref{a3}) then becomes
\ba
&&\frac{G_g(\ell)}{A} = 
k_{{\rm B}}T \, 
{\sum_{n=0}^{\infty}}' 
\!\int\! \frac{d^2\qv}{(2\pi)^2} 
\\
&&\times\ln \bigg(  
1 - \frac{u_1 u_3 + 8 \bar{\alpha}_1 \bar{\alpha}_3 + \bar{\alpha}_1^2 \bar{\alpha}_3^2}{v_1 v_3} \, {\rm e}^{-2q\ell} 
%\nonumber\\
%&&\qquad
+ \frac{\bar{\alpha}_1^2 \bar{\alpha}_3^2}{v_1 v_3} \, {\rm e}^{-4q\ell}
\bigg) 
\nonumber
\ea
We can check that this is the same result as that of Eq.~(\ref{G}) for isotropic topological insulators separated by a vacuum gap (i.e., $g_i=1$ ($i=1,2,3$), $\epsilon_{1z}=\epsilon_{3z}=\epsilon$ and $\epsilon_{2z}=1$).

\end{document}